\documentclass[prb, twocolumn, superscriptaddress, showpacs,floatfix]{revtex4-1}  
\usepackage{hyperref}
\usepackage{amssymb}
\usepackage{amsmath}
\usepackage{float}
\usepackage{bbm}
\usepackage{graphicx}
\usepackage{epsfig}
\usepackage{epstopdf}
\usepackage[usenames]{color}

\begin{document}
\bibliographystyle{apsrev4-1}

\title{Pairing Glue in the Two Dimensional Hubbard Model}

\author{E. Gull}
\affiliation{Department of Physics, University of Michigan, Ann Arbor, Michigan 48109, USA}

\author{A. J. Millis }
\affiliation{Department of Physics, Columbia University, New York, New  York 10027, USA}

\date{\today }

\begin{abstract}
Cluster dynamical mean field calculations are used to construct the  superconducting gap function of the two dimensional Hubbard model.   The frequency dependence of  the imaginary part of the gap function indicates that the pairing is dominated  by fluctuations at two characteristic frequencies: one at the scale of the hopping matrix element $t$ and one at a much lower scale.   The lower frequency component becomes more important as the doping is reduced into the pseudogap regime.  Comparison to available information on the spin fluctuation spectrum of the model suggests that  the superconductivity arises from exchange of spin fluctuations.  The inferred pairing glue function is in remarkable qualitative consistency with the pairing function inferred from time-resolved  optical conductivity data. \end{abstract}


\maketitle

The physical origin  and theoretical understanding of  the high transition temperature superconductivity observed \cite{Bednorz86} in layered copper oxide materials is an important open issue in condensed matter physics. One key question \cite{Anderson07,Maier08}  is  the extent to which superconductivity in these materials is due to fluctuations whose exchange provides a `pairing glue' binding electrons together into Cooper pairs. In conventional superconductors such as lead or mercury, superconductivity is generally believed to arise from exchange of phonons, collective fluctuations of ionic positions, whose properties and coupling to electrons are accurately described by Migdal-Eliashberg theory.\cite{Migdal58,Eliashberg60} In these conventional materials, direct evidence for the importance of phonons was obtained from theoretical \cite{Scalapino66} and experimental \cite{Rowell69} studies of the frequency dependent gap function, $\Delta(\omega)$, defined in more detail below. Within Migdal-Eliashberg theory,  $\Delta(\omega)$  has structure at the frequencies of the bosons making the dominant contribution to the superconducting pairing. It also has structure of the opposite sign at higher frequencies associated with the screened Coulomb interaction, which makes a repulsive contribution to the superconductivity.\cite{Scalapino66} Observation \cite{Rowell69} of these structures provided a definitive confirmation of the role of phonons in conventional superconductors.   

The superconductivity in  the copper-oxide high $T_c$ materials is believed to arise from electron-electron interactions, with phonons playing a minimal role. One of the central questions is whether the important effect of the interactions is to produce a collective electronic fluctuation (such as a magnon) whose exchange gives rise to superconductivity \cite{Monthoux92} or whether there is a pairing tendency intrinsic to strongly correlated materials in the vicinity of a  Mott state.\cite{Anderson07}  In situations where strong electron-electron interactions are dominant there is no a priori reason for  Migdal-Eliashberg equations to apply, although proximity to a quantum critical point may justify such a treatment in some cases.\cite{Millis92,Abanov01} However, it is plausible that even if a Migdal-Eliashberg treatment is not theoretically justified, the frequency dependence of $\Delta$ may provide insight into the origin of superconductivity,  with low frequency structure indicating fluctuation-mediated pairing while structure at high frequencies (for example on the order of the bare interaction strength)  might indicate a pairing tendency intrinsic to a strongly correlated Mott state.\cite{Anderson07,Maier08} 

Here we present results of a study of the $\Delta(\omega)$ corresponding to the d-wave superconducting state of the two dimensional Hubbard model,  a candidate model \cite{Anderson87} for the description of copper-oxide superconductivity.  Our results suggest that the superconductivity in this model is in fact driven by exchange of spin fluctuations, but reveal new features which remain to be understood. Our work is inspired in part by previous work of Maier and Scalapino \cite{Maier08} which aimed to extract  information about pairing from an analysis of the  anomalous self energy (not the gap function). Our results are not entirely consistent with this work.  We will explain the differences below.

The Hubbard model may be written in a mixed momentum/position representation as
\begin{align}
H=\sum_{k\sigma}c^\dagger_{k\sigma}\left(\varepsilon_k-\mu\right)c_{k\sigma}+U\sum_in_{i\uparrow}n_{i\downarrow},
\label{Hhub}
\end{align}
where the operator $c^\dagger_{k\sigma}$ creates an electron of spin $\sigma=\uparrow,\downarrow$ in momentum state $k$ and  $n_{i\sigma}$ is the operator giving the density of spin $\sigma$ electrons on site $i$.  We set the lattice constant to unity. In the case of interest here, the momentum index $k$ runs over the Brillouin zone of a two dimensional square lattice. The chemical potential is $\mu$ and $\varepsilon_k$ is the energy dispersion, which we take to have the simple nearest neighbor hopping form $\varepsilon_k=-2t\left(\cos k_x+\cos k_y\right)$.  The two dimensional square lattice Hubbard model is  known to exhibit both a `pseudogap' \cite{Gull10_clustercompare} and  $d_{x^2-y^2}$ superconductivity.\cite{Zanchi96,Halboth00,Maier05_dwave,Raghu10} Many properties of the superconducting state including the doping dependence of the superconducting phase diagram \cite{Gull12} and the interplay of the  photoemission,\cite{Gull12} Raman and interplane conductivity spectra \cite{Gull13} with the pseudogap have been shown to be in good qualitative agreement with experiment.  However the physical origin of the superconductivity has remained unclear.

The first issue in our study is the  definition of $\Delta$. The electron Green function in the superconducting state may be written on the Matsubara axis as
\begin{eqnarray}
&&\mathbf{G}(k,\omega_n)^{-1}=
\label{Gdef}
\\
&&\hspace{0.1in}\left(\begin{array}{cc}i\omega_n-\varepsilon_k-\Sigma^N(k,\omega_n) & \Sigma^A(k,\omega) \\\Sigma^A(k,\omega) &i\omega_n+\varepsilon_k+\Sigma^{N}(k,-\omega_n) \end{array}\right)
\nonumber
\end{eqnarray}
where $\Sigma^{N,A}$ are the normal and anomalous components of the electron self energy and we have chosen phases so that the anomalous self energy is real.  Eq.~\ref{Gdef} implies that 
\begin{eqnarray}
det\left(G^{-1}\right)&=&-\left(1-\frac{\Sigma^N_o(k,\omega_n)}{i\omega_n}\right)
\label{detG} \\
&\times&\left(\omega_n^2+\left(\varepsilon^\star(k,\omega_n)\right)^2+\Delta^2(k,\omega_n)\right)
\end{eqnarray}
with 
\begin{eqnarray}
\Sigma^N_{o,e}&=&\frac{\Sigma^N(k,\omega_n)\mp\Sigma^N(k,-\omega_n)}{2}
\\
\varepsilon^\star&=&\frac{\varepsilon_k+\Sigma^N_e(k,\omega_n)}{1-\frac{\Sigma^N_o(k,\omega_n)}{i\omega_n}}
\\
\Delta(k,\omega_n)&=&\frac{\Sigma^A(k\omega_n)}{1-\frac{\Sigma^N_o(k,\omega_n)}{i\omega_n}}
\label{Deltadef}
\end{eqnarray}
We identify the Fermi surface (renormalized by interactions and possibly changed by superconductivity) as the locus of k-points such that $\varepsilon^\star(k,\omega=0)=0$ so that $\Delta$ is the gap at the Fermi surface. IN Migdal-Eliashberg theory the $\Delta$ defined in this way has structure at the frequencies of the pairing phonons. We propose that $\Delta$ contains information about pairing more generally. 
To calculate $\Delta$ we use the DCA (`dynamical cluster approximation')  version \cite{Hettler98,Hettler00} of cluster dynamical mean field theory \cite{Maier05} along with the continuous-time auxiliary field (CT-AUX)\cite{Gull08} implementation of the continuous-time quantum Monte Carlo algorithm \cite{Rubtsov05,Gull11} and submatrix updates.\cite{Gull10_submatrix} In the DCA  the Brillouin zone is partitioned into $N$ equal area tiles labeled by central momentum $K$ and the self energy  is approximated as a piecewise continuous function
\begin{equation}
\mathbf{\Sigma}(k,\omega)=\sum_K^N\phi_K(k)\mathbf{\Sigma}_K(\omega)
\label{DCA}
\end{equation}
with $\phi_K(k)=1$ if $k$ is in the tile centered on $K$ and 0 otherwise. The self energies $\mathbf{\Sigma}$ are matrices in Nambu space with normal and anomalous components and are obtained from the solution of an auxiliary quantum impurity model. $\Delta$ is constructed as a function of Matsubara frequencies from the self energies via Eq.~\ref{Deltadef}.

\begin{figure}[t]
\includegraphics[width=0.9\columnwidth]{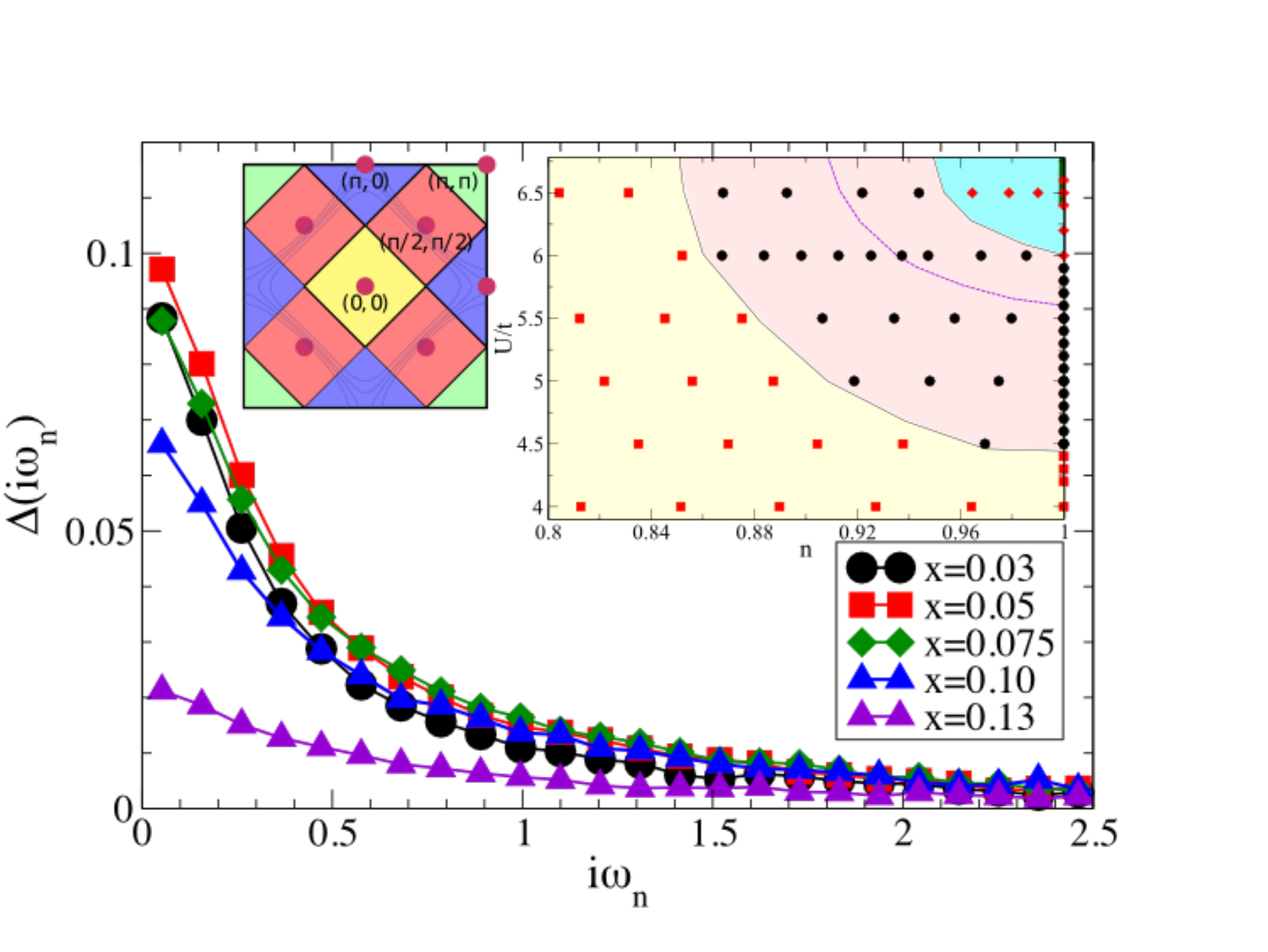}
\caption{$\Delta(i\omega_n)$ as a function of Matsubara frequency for different dopings at $U=6t$. Left inset: momentum space tiling of the DCA cluster used in this publication. Right inset: phase diagram according to Ref.~\onlinecite{Gull12b}.  }
\label{fig:deltamat}
\end{figure}

The expense of the computation increases rapidly with increasing interaction strength, increasing number of approximants $N$ and decreasing temperature. We present results for interaction strength $U=6t$  using $N=8$ approximants with the standard  momentum-space tiling (see left inset to Fig.~\ref{fig:deltamat}).  Previous work \cite{Gull10_clustercompare,Gull12,Gull13} has shown that  $N=8$ is large enough to be representative of the $N\rightarrow\infty$ limit, being in particular  large enough to represent the difference between zone-diagonal and zone-face electronic properties and therefore large enough to capture the essential physics including a paramagnetic insulating phase at carrier concentration $n=1$ per site, a pseudogap regime and $d_{x^2-y^2}$-symmetry superconductivity existing within a superconducting dome (see phase diagram in right inset of Fig.~\ref{fig:deltamat}). Comparison of results for various physical quantities including the magnitude of the pseudogap and the density of the pseudogap onset calculated for different cluster sizes suggests quantitative accuracy on the $\sim 25\%$ level.\cite{Gull10_clustercompare} The value $U=6t$ was chosen to be small enough to permit  calculations in the superconducting phase with the precision needed for reliable analytical continuation of self energies and gap functions, yet large enough to capture the essential physics. However, for $U=6t$ the superconducting dome is pushed closer to half filling than is the case in actual materials.   The 8 square tiles are the zone center and zone corner  momentum sectors $K=(0,0),(\pi,\pi)$, the four symmetry-equivalent zone diagonal sectors centered on $K=(\pm \pi/2,\pm\pi/2)$ and the two zone-face sectors $K=(\pi,0)$ and $(0,\pi)$. Note that in the $N=8$ d-wave state  symmetry considerations imply that the anomalous self energy is only non-zero in the sectors centered on $(0,\pi)$ and $(\pi,0)$ and $\Sigma^A_{K=(\pi,0)}(\omega)=-\Sigma^A_{K=(0,\pi)}(\omega)\equiv \Sigma^A(\omega)$.  We focus on this sector in what follows, and suppress the explicit momentum arguments. 

The CT-AUX method yields results on the imaginary (Matsubara) frequency axis. The main panel of Fig.~\ref{fig:deltamat} shows the doping dependence of the resulting Matsubara-axis gap function for five four dopings spanning the superconducting region of the phase diagram. We see that in all cases the gap function drops rapidly with frequency, becoming indistinguishable from 0 (within our error bars) for Matsubara frequencies greater than about $2.5t$. We also see that the gap function is weakly doping dependent in the middle of the superconducting region, but drops as the edge of the superconducting dome is reached on the high doping side. A similar drop in $\Delta$ occurs on the low doping side of the superconducting dome. The start of this drop may be seen in the $x=0.03$ data. 

We now turn to the behavior of $\Delta$ on the real frequency axis. Viewed as a function of complex variable $z$, $\Delta(z)$ is analytic in the complex plane except for a branch cut along the real frequency axis,  \cal{Im}~$z=0$. It  has a spectral representation 
\begin{equation}
\Delta(z)=\int\frac{dx}{\pi}\frac{ \text{\cal{Im}}\Delta(x)}{z-x}.
\label{Deltakk}
\end{equation}
The spectral function $\text{\cal{Im}}\Delta(x)$ is the object of primary physical interest in the Migdal-Eliashberg-Scalapino-Rowell analysis.\cite{Scalapino66,Rowell69}   Direct inversion of Eq.~\ref{Deltakk} to find $\text{\cal{Im}}\Delta$ in terms of the computed quantity  $\Delta(i\omega_n)$ is a mathematically ill-posed problem, necessitating use of a numerical analytical continuation process.\cite{Jarrell96} The Matsubara axis $\Delta$ is an even function of frequency, implying that the spectral function $\text{\cal{Im}}\Delta$ is an odd function of frequency whereas the standard maximum entropy continuation methodology \cite{Jarrell96} requires a non-negative spectral function.  We therefore rearrange Eq.~\ref{Deltakk} as 
\begin{eqnarray}
\Delta(i\omega_n)=\Delta(i\omega_n=0)+i\omega_n\int \frac{dx}{\pi}\frac{\frac{\Delta^{(2)}(x)}{x}}{i\omega_n-x}
\end{eqnarray} 
and continue $\frac{\Delta(i\omega_n)-\Delta(i\omega_n=0)}{i\omega_n}$ by standard methods. We obtain $\Delta(i\omega_n=0)$ by fitting $\Delta(i\omega_{n=1,2,3})$ at the lowest three Matsubara frequencies to a parabola. 

A potentially serious difficulty  is that  there no guarantee that  $\text{\cal{Im}}\Delta/\omega$ (or equivalently $\Sigma^{A,(2)}/\omega$) is of definite sign. For example, in the usual Migdal-Eliashberg theory the Coulomb pseudopotential leads to a sign change at frequencies somewhat above the phonon frequencies, reflecting the repulsive (depairing) contribution of the Coulomb repulsion in conventional metals.\cite{Scalapino66}  The results presented in Ref.~\onlinecite{Maier08} are  consistent with a weakly negative $\Sigma^A$ in certain frequency regimes. A recent solution of the Eliashberg equations for a model involving two competing spin fluctuations also displayed a sign change in the gap function as frequency was increased above a characteristic frequency.\cite{Fernandes13} 

We have investigated the sign of $\text{\cal{Im}}\Delta(\omega)/\omega$ in two ways. First, we crosschecked our results by use of a Pad\'{e} continuation method \cite{Beach00} that makes no assumption about the sign of $\text{\cal{Im}}\Delta(\omega)/\omega$. This method consistently found a positive-definite $\text{\cal{Im}}\Delta(\omega)/\omega$ with no evidence for any sign change. Second, we considered the  particle-hole symmetric ($n=1$) situation. In this case the self energy for the $(\pi,0)$ sector is also particle-hole symmetric and obeys the condition $\Sigma^N(z)=-\Sigma^{N,*}(-z)$ so that the impurity model Green function (in the $K=(0,\pi)$ sector) and the self energy matrix  are diagonalized at all frequencies by the Majorana combinations $c^\dagger_{k\sigma}\pm c_{-k,-\sigma}$. In this $\pm$ basis we have
\begin{equation}
\Sigma^\pm(z)=\Sigma^N(z)\pm\Sigma^A(z).
\label{pmdef}
\end{equation}
Because the Greens function and  $\Sigma$ are diagonal in the $\pm$ basis  the associated spectral functions are positive definite so standard maximum entropy methods may be used. We have constructed $\text{Im}\Delta$  in the $\pm$ basis for a range of $U$ at $n=1$, finding results in agreement with direct continuations of $\Delta$ obtained on the assumption that the spectral function associated with $\Sigma^A$ is non-negative. 

\begin{figure}[t]
\includegraphics[width=0.95\columnwidth]{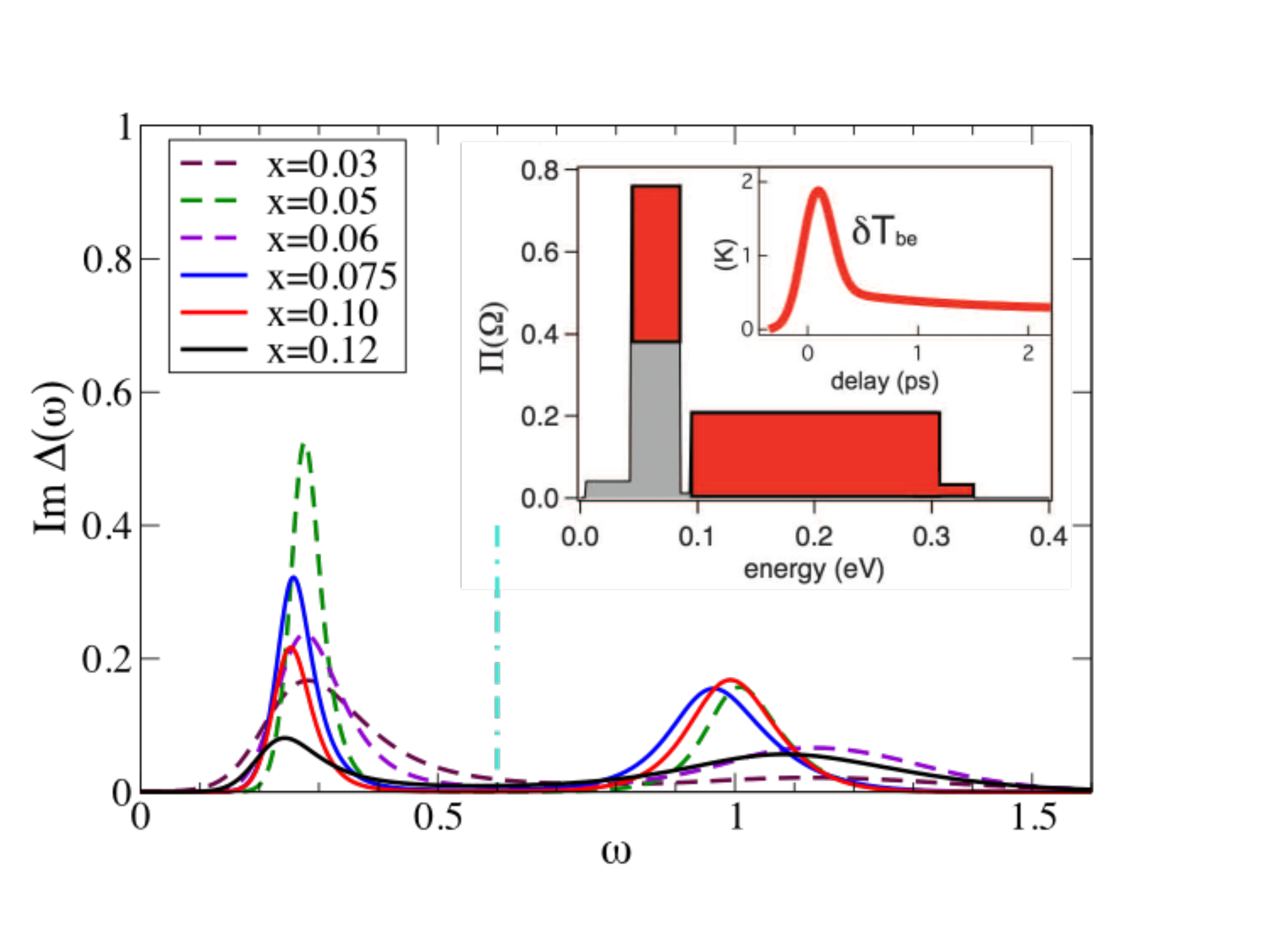}\caption{Main panel: Imaginary part of real frequency gap function computed for different dopings. Dashed curves label dopings within the pseudogap regime and solid curves to dopings outside the pseudogap regime. Vertical dashed line indicates the frequency cutoff chosen in the inset of Fig.~\ref{fig:imdeltadopingnewinset}. Inset: Experimental data reproduced from Ref.~\onlinecite{DalConte12}.}
\label{fig:imdeltadoping}
\end{figure}

Fig.~\ref{fig:imdeltadoping} shows our principal results: the imaginary part of the gap function of the Hubbard model, computed for different dopings in the superconducting regime of the phase diagram. The support for the spectral function is concentrated in two regions: a peak at the very low frequency $0.25t\lesssim 0.1eV$ (with the usual identification $t\sim 0.3eV$ for cuprates)  and a higher peak at a frequency $\sim t$.  This two-peak structure is robustly found in continuations of all of our superconducting state data and although the method is subject to non-negligible systematic uncertainties especially at higher frequencies, the crucial aspects of the results  can be inferred directly from the Matsubara axis data.  

The first important qualitative result is that $\text{\cal{Im}}\Delta$ has negligible support at frequencies higher than those shown in Fig.~\ref{fig:imdeltadoping}. This result is confirmed by the rapid decrease of $\Delta$ with increasing Matsubara frequency   displayed in Fig.~\ref{fig:deltamat}.  If $\text{\cal{Im}}\Delta$ had significant support at higher frequencies, $\Delta(i\omega_n)$ would not decrease so rapidly to zero. In particular, Ref.~\onlinecite{Maier08} reported results of a study of the $N=4$ approximation using an `NCA' impurity solver that about $20\%$ of the pairing came from much higher frequencies, of the order of $U$. If this were the case, $\Delta(i\omega_n\approx 2t)$ would be about $20\%$ of its value at $\omega_n=\pi T$. Our Matsubara axis data clearly rule out this possibility, and an independent analysis of the $N=4$ approximation by Civelli (cf Fig.~5 in \textcite{Civelli09}) also found that $\text{Im}\Sigma^A$ goes rapidly to $0$ for frequencies above $\sim 1.5t$.  The difference may arise from the use of the NCA solver in Ref.~\onlinecite{Maier08}.   We conclude that in the Hubbard model,  pairing comes from frequencies at most of order $\sim t$, well below the energy of the upper Hubbard band.

We now turn to the detailed frequency dependence. The existence of a very  low-frequency peak ($\omega\sim0.25t\sim 75meV$, using the $t\approx0.3eV$ appropriate to cuprates)  in $\text{\cal{Im}}\Delta$ is a surprising feature of our results. We believe that it is not an artifact of the   maximum entropy analytical continuation method used here. This method is generally found to yield reliable results for the lowest frequency features. We have confirmed the results by performing Pad\'{e} continuations (not shown), which reproduce the position and spectral weight of the low frequency peak.    The existence of significant spectral weight at higher frequencies is also directly implied by the Matsubara axis data:  analysis (not shown)  of the Matsubara axis data in Fig.~\ref{fig:deltamat} reveals that  $\Delta(i\omega_n)$ decays more slowly than $\omega_n^{-2}$ for $\omega_n\lesssim t$ contradicting the hypothesis  that the $\omega\approx0.25t$ feature is the only structure in $\text{\cal{Im}}\Delta$.  Our experience is that for higher frequency features maximum entropy analytical continuation provides reasonable estimates for spectral weights in given frequency regimes, but is not necessarily reliable for precise position and shape of spectral features. Thus we believe that while the existence of a low frequency peak and a higher frequency  structure in $\text{\cal{Im}}\Delta$  as well as their relative weights are clearly established, the structure of two sharp peaks indicated by the analytical continuation is not yet proven.

\begin{figure}[t]
\includegraphics[width=0.95\columnwidth]{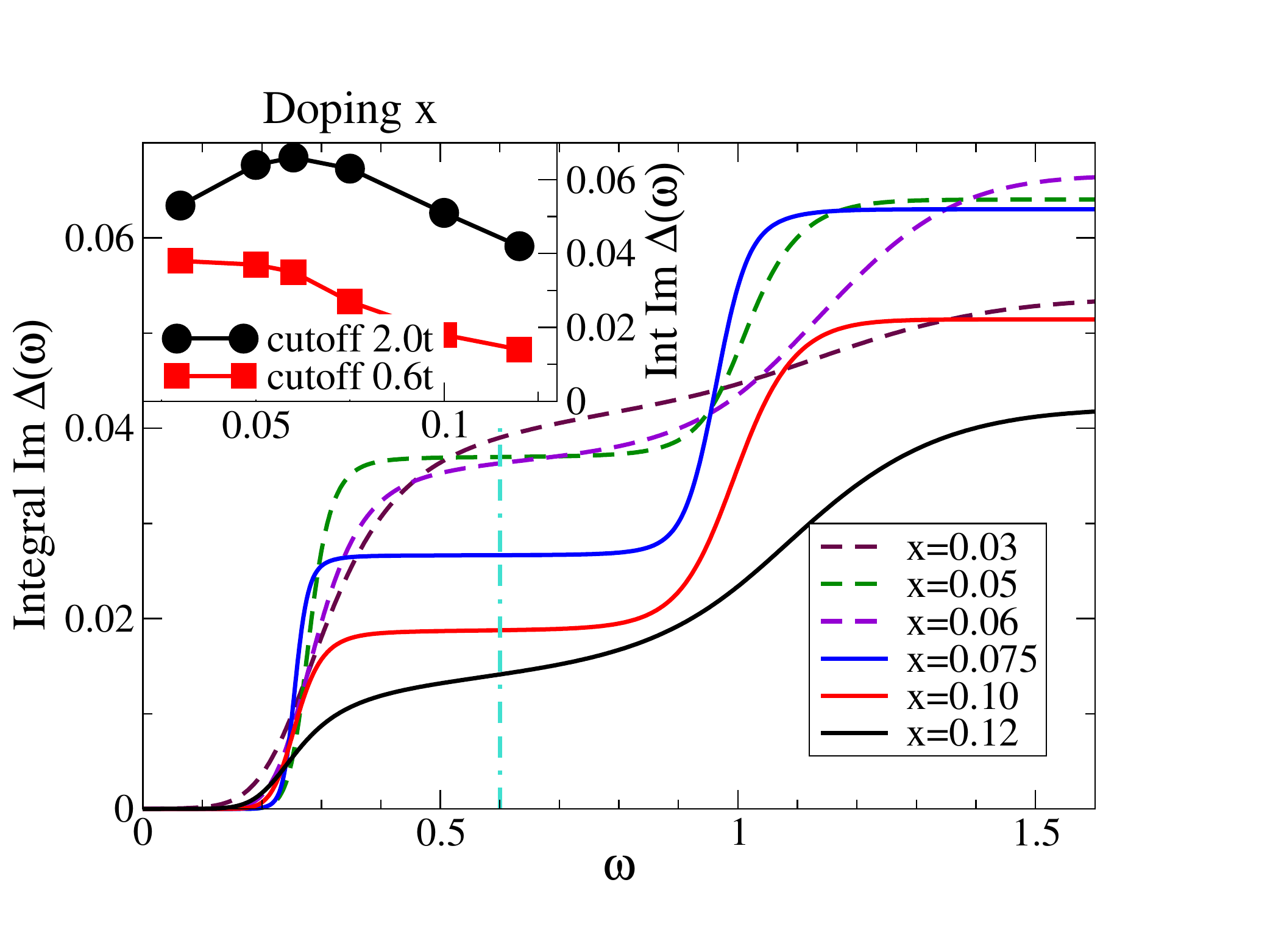}\caption{
Main panel: Partial integral of $\text{Im}\Delta(\omega)$ for the dopings of Fig.~\ref{fig:deltamat} up to $\Omega=1.6t$. Vertical dashed line: cutoff frequency used in inset. Inset: integral over the entire frequency range and over range of low frequency pole, as a function of doping.}
\label{fig:imdeltadopingnewinset}
\end{figure}

That Im $\Delta$ is non-negligible only at frequencies  $\omega \lesssim t \approx 0.3eV$ which are low compared to the intrinsic scales of the model such as bandwidth suggests that the superconductivity arises from exchange of a relatively low frequency collective electronic excitation.  We observe that in the Hubbard model at these interaction scales, the basic spin fluctuation energy (zone boundary magnon frequency)  $\omega_{SF}$ is of the order of $t$: this may be seen from Fig.~1 of Ref.~\onlinecite{Lin12} (note that the two-magnon peak in the Raman scattering occurs at about the same energy as the maximum of the single-magnon energy); see also Fig.~3 of Ref.~\onlinecite{Maier08}. This suggests, in agreement with the results of Ref.~\onlinecite{Maier08}, that the pairing is spin fluctuation-driven.

The sharp low frequency peak  is remarkable. The  inset of Fig.~\ref{fig:imdeltadoping} which compares $\int_0^{0.6t}\text{\cal{Im}}\Delta(\omega)d\omega$ to $\int_0^{2.0t}\text{\cal{Im}}\Delta(\omega)d\omega$, shows that the relative importance of the low frequency feature increases as doping is decreased into the pseudogap regime. The peak position is seen to be approximately the same for all dopings, whereas Fig.~\ref{fig:deltamat} shows that the gap value $\approx \lim_{\omega\rightarrow 0}\Delta(i\omega_n)$ varies substantially with doping. We therefore believe that although for intermediate dopings the peak energy is approximately three times the gap value, the peak feature  is not simply an above-gap excitation, but corresponds to a physically significant fluctuation, related in some way to the pseudogap.  In  remarkable recent experiments,\cite{Heumen09,DalConte12}  optical measurements  were used to infer a pairing glue spectrum consisting of a sharp peak centered at $\omega\approx 0.07eV$ and a broad continuum extending up to $\approx 0.3eV$, in striking agreement with the numerical results presented here. Further investigation of the physics of this structure is an important open problem.

In conclusion, we have revealed  insights into the superconducting state of the two dimensional Hubbard model. A definition of the gap function valid beyond the Migdal-Eliashberg approximation was introduced, and structure in this gap function was found to indicate that superconductivity arises from exchange of relatively low frequency collective electronic fluctuations, presumably of magnetic origin. However the gap function exhibits an unanticipated  very low frequency ($\sim 0.25t$) feature  of unknown origin. Understanding the physics of this feature is an important open question. We also remark that neither the Matsubara-axis nor the continued data provides evidence for  the power-law scaling of $\Delta$ predicted by  quantum critical theories  of strongly correlated superconductivity.\cite{Abanov01}

Acknowledgements: We thank D. van der Marel for helpful comments on the manuscript and D. Scalapino for correcting an error in the definition of $\Delta$. The research was supported by NSF-DMR-1308236 (A.J.M.) and the Sloan foundation (E.G.). A portion of this research was conducted at the National Energy Research Scientific Computing Center (DE-AC02-05CH11231), which is supported by the Office of Science of the U.S. Department of Energy. Our continuous-time quantum Monte Carlo codes are based on the ALPS\cite{ALPS20,ALPS_DMFT} libraries.

\bibliography{refs_shortened.bib}
\end{document}